\documentclass{ws-procs9x6}
\usepackage{subfigure}

\begin{document}

\title{Observation of extragalactic sources of very high energy $\gamma$-rays with the MAGIC telescope}

\author{M. Errando for the MAGIC Collaboration\footnote{Updated collaborator list at http://wwmagic.mppmu.mpg.de/collaboration/members}}

\address{Institut de Física d'Altes Energies (IFAE)\\
Edifici Cn., Universitat Autònoma de Barcelona\\
08193 Bellaterra (Spain)\\
E-mail: errando@ifae.es}

\begin{abstract}
MAGIC is currently the largest single dish ground--based imaging air Cherenkov telescope in operation. During its first cycle of observations more than 20 extragalactic objects have been observed, and very high energy $\gamma$-ray signals have been detected in several of them. The results of this observations are presented, together with a discussion of the spectral characteristics and the flux variability of the detected sources.

\end{abstract}

\section{The MAGIC telescope}
MAGIC is a 17m imaging atmospheric Cherenkov telescope for Very High Energy (VHE) $\gamma$-ray observations. It is located in the Roque de los Muchachos Observatory on the canary island of La Palma. It can explore gamma-rays at energies below 100\:GeV, which is critical to observe sources at medium redshifts where absorption of gamma photons by the Extragalactic Background Light (EBL) attenuates the emission at higher gamma energies.

The detection and characterization of VHE $\gamma$-ray emitting Active Galactic Nuclei (AGN) is one of the main goals for ground--based $\gamma$-ray astronomy. This studies open the possibility of exploring the physics of the relativistic jets in AGN, relate the flux of photons in different energy bands (optic, X-rays and $\gamma$-rays), perform population studies of blazar objects and even extract information about the EBL photon density. MAGIC can also be repositioned in few seconds to do observations of Gamma-Ray Bursts (GRBs) in their prompt phase of early afterglow, although none of these has resulted in a positive detection to date. GRB observations done by MAGIC are reviewed elsewhere in these proceedings\cite{grb}, as well as the results of galactic observations\cite{galactic}.


\section{Cycle I observations}
The first cycle of observations of the MAGIC telescope extended from January 2005 to April 2006. The observations of extragalactic objects covered High-frequency peaked BL Lacs (HBLs), selected flat spectrum radio quasars and low-frequency peaked BL Lacs located at low redshifts, known TeV-emitting HBLs, the Ultra--Luminous infrared Galaxy (ULIRG) \mbox{Arp 220} and the radio galaxy M87. In this section, the following results are reviewed: the recently discovered VHE $\gamma$-ray emission from the HBLs \mbox{1ES 1218+30.4}, Mkn 180 and PG 1553+113, an upper limit on the VHE emission of Arp 220 and the observations of the known TeV blazars \mbox{Mkn 501}, Mkn 421, 1ES 1959+650 and 1ES 2344+514.

\begin{figure}[t]
	\centering
	\includegraphics[height=5cm,bb=14 14 582 398]{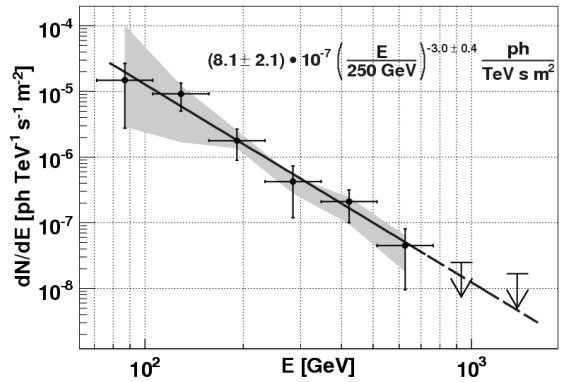}
	\caption{Differential energy spectrum of 1ES 1218+30.4.}
	\label{fig:1218}
\end{figure}

\subsection{Observation of VHE $\gamma$-ray candidate sources}
Selection of VHE $\gamma$-ray emitting candidate sources follows criteria based on the spectral properties of the considered objects. Using both Synchrotron Self--Compton (SSC) and hadronic models, the spectral energy distribution of the candidate AGN can be extrapolated to MAGIC energies to predict its observability. The preferred candidates are usually strong X-ray emitters, but selections based on the optical band are also considered. Moreover, other non--blazar objects as the giant radio--galaxy M87 or the ULIRG \mbox{Arp 220} have also been observed, although none of these observations resulted in a positive detection to date.\\


\textbf{1ES 1218+30.4}\cite{1218} is the first source discovered by MAGIC and one of the most distant VHE $\gamma$-ray sources. This HBL located at redshift $z=0.182$ was previously observed by Whipple and HEGRA, but only upper limits where derived. MAGIC observed it during 8.2\:h in January 2005, obtaining a $\gamma$-ray signal of 6.4\:$\sigma$ significance in the 87 to 630 GeV energy range. The observed differential energy spectrum can be fitted by a simple power law with photon index $3.0\pm0.4$, and no time variability on timescales of days was found within statistical errors.\\

\begin{figure}[t]
	\centering
	\includegraphics[height=5cm,bb=14 14 505 349]{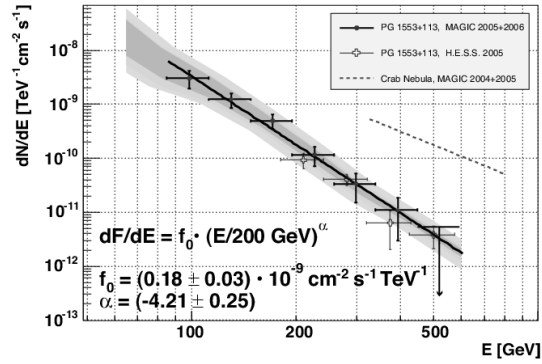}
	\caption{Differential energy spectrum of PG 1553+113.}
	\label{fig:1553}
\end{figure}

\textbf{PG 1553+113}\cite{1553} is a distant BL Lac of undetermined redshift that was observed by the MAGIC telescope in 2005 and 2006, and has also been recently detected by the H.E.S.S. collaboration\cite{hess1553}. A VHE $\gamma$-ray signal has been detected by MAGIC with an overall significance of 8.8\:$\sigma$, showing no significant flux variations a daily timescale. However, the flux observed in 2005 was significantly higher compared to 2006. The differential energy spectrum between 90 and 500\:GeV can be well described by a power law with photon index $4.2\pm0.3$, being steeper than that of any other known BL Lac object. This spectrum can be used to derive an upper limit on the source redshift. Assuming an EBL model by Kneiske et al.\cite{kneiske} and a physical limit on the intrinsic source spectrum (intrinsic photon index $\alpha_{int}<-1.5$)\cite{limit}, an upper limit on the source redshift of $z<0.78$ has been derived.\\


\begin{figure}[t]
	\centering
	\includegraphics[height=5cm,bb=0 0 1465 858]{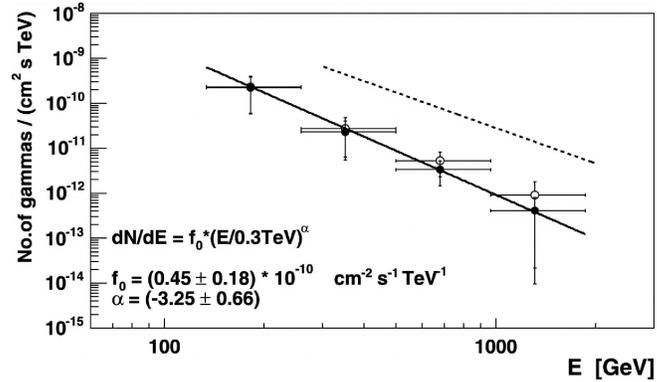}
	\caption{Differential energy spectrum of Mkn 180. The filled circles show the measured spectrum while the open circles display the intrinsic spectrum of the source where the EBL absorption has been removed.}
	\label{fig:180}
\end{figure}

\textbf{Mkn 180}\cite{180} is an HBL that had an optical outburst in March 2006 observed by the KVA 35\:cm telescope also located at the Roque de los Muchachos Observatory. It triggered the observations with the MAGIC telescope in the GeV--TeV band, resulting on the first detection of VHE $\gamma$-ray emission from this source. A total of 12.4\:h of data were recorded during eight nights, giving a 5.5\:$\sigma$ significance detection. The integral flux above 200\:GeV corresponded to 11\% of the Crab Nebula flux, and the differential energy spectrum could be fitted by a power law with a photon index of $3.3\pm0.7$.\\

\textbf{Arp 220}\cite{220} is the nearest ULIRG (located at about 72\:Mpc) and the one with the largest supernova explosion rate ($4\pm2$ per year), and therefore is a good VHE $\gamma$-ray emitting candidate. With 15.5\:h of data taken by MAGIC upper limits in the 0.16-1.3 TeV band were imposed, which are compatible with a complete multiwavelength modeling of the source \cite{torres}.

\section{Monitor of known TeV blazars}
The improved sensitivity and energy threshold of MAGIC with respect to the former generation of $\gamma$-ray telescopes allows a detailed study of the spectral features and flux variations of known TeV emitters.\\

\textbf{Mkn 421}\cite{421} is the closest TeV blazar ($z=0.031$) and the first extragalactic VHE source detected with a ground--based $\gamma$-ray telescope\cite{punch}. MAGIC has observed this source between November 2004 and April 2005 obtaining 25.6\:h of data, and including 1.5\:h of simultaneous observations with the H.E.S.S. array \cite{mazin}. Integral flux variations up to a factor of four are observed between different observation nights, although no significant intra--night variations have been recorded despite the high sensitivity of the MAGIC telescope for this kind of search. This flux variability showed a clear correlation with between $\gamma$-ray and X-ray fluxes, favoring leptonic emission models. The energy spectrum between 100\:GeV and 3\:TeV shows a clear curvature. After correcting the measured spectrum for the effect of $\gamma$-attenuation caused by the EBL light assuming a model of Primack et al.\cite{primack}, there is an indication of an inverse Compton peak around 100\:GeV.\\

\textbf{1ES 2344+514} was first detected in 1995 by Whipple when it was in flaring state\cite{w2344}, and HEGRA reported later a weak detection in quiescent state\cite{h2344}. MAGIC obtained a VHE $\gamma$-ray signal with 11.0\:$\sigma$ significance from 23.1\:h of data. The source was in quiescent state during the observations, with a flux level compatible with the HEGRA results, but showing a softer spectrum. A detailed publication on the analysis and results of these observations is in preparation.\\

\textbf{1ES 1959+650}\cite{1959} presented in 2002 a VHE $\gamma$-ray flare without any counterpart in X-rays \cite{kraw}. This behavior can not be easily explained by the SSC mechanism in relativistic jets that successfully explain most of the VHE $\gamma$-ray production in other HBLs. MAGIC observed this object during 6\:h in 2004, when it was in low activity both in optical and X-ray bands, detecting a $\gamma$-ray signal with 8.2\:$\sigma$ significance. The differential energy spectrum between 180\:GeV and 2\:TeV can be fitted with a power law of photon index $2.72\pm0.14$, which is consistent with the slightly steeper spectrum seen by HEGRA at higher energies\cite{h1959}, also during periods of low X-ray activity.\\

\textbf{Mkn 501} is a close TeV blazar, first detected by Whipple in 1996\cite{w501}. MAGIC observed it during 55\:h in 2005, including 34\:h in moderate moonlight conditions. The source was in low state (30-50\% of the Crab Nebula flux for $E>200$\:GeV) during most of the observation time but showed two episodes of fast and intense flux variability, with doubling times of about 5 minutes. Changes in the spectral slope with the flux level have been observed for the first time in timescales of 10 minutes. A detailed publication on the analysis and results of these observations is in preparation.

\section{The MAGIC II telescope}
The MAGIC collaboration is currently constructing a second telescope in the same site at the Roque de los Muchachos Observatory, which will operate in stereo mode with the MAGIC telescope improving the overall sensitivity. The MAGIC II telescope is a clone of the existing one with two main improvements: a fine pixelized camera with cluster design that will allow an update of the photomultipliers to hybrid photon detectors once this technology is ready to be used, and an ultra--fast signal readout that will sample the pulses coming from each pixel at 2.5\:GHz, allowing a better signal reconstruction and the use of timing analysis techniques. The commissioning phase of MAGIC II will start at the end of 2007.

\section{Conclusions}
MAGIC has concluded its first cycle of observations having detected seven extragalactic VHE $\gamma$-ray sources, including three objects never detected before at these wavelengths. The high sensitivity and low energy threshold allowed detailed studies of the spectral features of these sources, as well as the observation of flux variability in short timescales.

\end{document}